\PassOptionsToPackage{unicode}{hyperref}
\PassOptionsToPackage{hyphens}{url}
\PassOptionsToPackage{dvipsnames,svgnames,x11names}{xcolor}
\documentclass[12pt]{article}
\usepackage{amsmath,amssymb}
\usepackage{lmodern}
\usepackage{iftex}
\ifPDFTeX
  \usepackage[T1]{fontenc}
  \usepackage[utf8]{inputenc}
  \usepackage{textcomp} 
\else 
  \usepackage{unicode-math}
  \defaultfontfeatures{Scale=MatchLowercase}
  \defaultfontfeatures[\rmfamily]{Ligatures=TeX,Scale=1}
\fi
\IfFileExists{upquote.sty}{\usepackage{upquote}}{}
\IfFileExists{microtype.sty}{
  \usepackage[]{microtype}
  \UseMicrotypeSet[protrusion]{basicmath} 
}{}
\makeatletter
\@ifundefined{KOMAClassName}{
  \IfFileExists{parskip.sty}{%
    \usepackage{parskip}
  }{
    \setlength{\parindent}{0pt}
    \setlength{\parskip}{6pt plus 2pt minus 1pt}}
}{
  \KOMAoptions{parskip=half}}
\makeatother
\usepackage{xcolor}
\usepackage{graphicx}
\makeatletter
\def\maxwidth{\ifdim\Gin@nat@width>\linewidth\linewidth\else\Gin@nat@width\fi}
\def\maxheight{\ifdim\Gin@nat@height>\textheight\textheight\else\Gin@nat@height\fi}
\makeatother
\setkeys{Gin}{width=\maxwidth,height=\maxheight,keepaspectratio}
\makeatletter
\def\fps@figure{htbp}
\makeatother
\NewDocumentCommand\citeproctext{}{}
\NewDocumentCommand\citeproc{mm}{%
  \begingroup\def\citeproctext{#2}\cite{#1}\endgroup}
\makeatletter
 \let\@cite@ofmt\@firstofone
 \def\@biblabel#1{}
 \def\@cite#1#2{{#1\if@tempswa , #2\fi}}
\makeatother
\newlength{\cslhangindent}
\newlength{\csllabelwidth}
\newenvironment{CSLReferences}[2] 
 {\begin{list}{}{%
  \setlength{\itemindent}{0pt}
  \setlength{\leftmargin}{0pt}
  \setlength{\parsep}{0pt}
  \ifodd #1
   \setlength{\leftmargin}{\cslhangindent}
   \setlength{\itemindent}{-1\cslhangindent}
  \fi
  \setlength{\itemsep}{#2\baselineskip}}}
 {\end{list}}
\usepackage{calc}

\ifLuaTeX
\usepackage[bidi=basic]{babel}
\else
\usepackage[bidi=default]{babel}
\fi
\babelprovide[main,import]{american}

\def\languageshorthands#1{}
\ifLuaTeX
  \usepackage{selnolig}  
\fi
\IfFileExists{bookmark.sty}{\usepackage{bookmark}}{\usepackage{hyperref}}
\IfFileExists{xurl.sty}{\usepackage{xurl}}{} 
\hypersetup{
  pdftitle={commensurability: a Python package for classifying
astronomical orbits based on their toroid volume},
  pdfauthor={Subhadeep Sarkar, Michael S. Petersen},
  pdflang={en-US},
  colorlinks=true,
  linkcolor={Maroon},
  filecolor={Maroon},
  citecolor={Blue},
  urlcolor={Blue},
  pdfcreator={LaTeX via pandoc}}

\title{\texttt{commensurability}: a Python package for classifying astronomical
orbits based on their toroid volume}

\usepackage[affil-it]{authblk}
\usepackage{orcidlink}
\author[1,2%
  \ensuremath\mathparagraph]{Subhadeep Sarkar%
    \,\orcidlink{0009-0006-9454-5141}\,%
    }
\author[2%
  ]{Michael S. Petersen%
    \,\orcidlink{0000-0003-1517-3935}\,%
    }

\affil[1]{Institut de Ciències del Cosmos, Universitat de Barcelona,
Spain%
  }
\affil[2]{Institute for Astronomy, University of Edinburgh, UK%
  }
\affil[$\mathparagraph$]{Corresponding author%
}
\date{\today{}}

\usepackage[margin=1.05in]{geometry}
\usepackage[scaled]{beramono}

\definecolor{darkgreen}{rgb}{0.05, 0.3, 0.1}
\let\oldtexttt\texttt
\renewcommand{\texttt}[1]{\oldtexttt{\textcolor{darkgreen}{#1}}}
\usepackage{caption}

\begin{document}
\maketitle


\section{Summary}\label{summary}

Stars like our Sun orbit the center of our Milky Way galaxy. Over many
orbits in the stellar disk of the galaxy, a star will (hypothetically)
probe every point in a 3D toroid defined by a minimum/maximum radius and
minimum/maximum height above the midplane. However, there are special
classes of orbits that repeat their tracks over many revolutions around
the Galactic center, and therefore only probe a small sub-volume of the
toroid. This property stems from low-integer orbital frequency
ratios---commensurate frequencies---and such orbits are referred to as
satisfying a ``commensurability''.

To study the orbital content of galaxies, astronomers often use models
for galaxies that allow for the integration of orbits within the model
galaxy's potential. The integration results in a time series of
positions (e.g.~\(x,y,z\)) for the orbit that can then be analyzed to
produce a classification. Complicating this picture, potentials evolving
with time introduce a new frequency: the pattern frequency of the
evolution. The evolution of the potential might be sourced by the
rotation of a galactic bar, the formation or dissolution of spiral arms,
the growth of the galaxy, or the interaction with a satellite galaxy.
Orbits that are commensurate with the pattern frequency are particularly
important in galactic dynamics as they experience the same force
fluctuations during every revolution: this causes changes to the
typically conserved quantities of orbits (e.g.~energy and angular
momentum).

Given their importance, astronomers have developed techniques to
identify commensurate orbits. One technique involves tessellating
orbital coordinates to estimate the (sub-)volume of the toroid traversed
(\citeproc{ref-Petersen.Weinberg.Katz.2021}{Petersen et al., 2021}).
This ``orbit tessellation'' method aims to pick out commensurate orbits
over relatively short orbit integration times, as well as pick out
commensurabilities in cases where the kinematic frequencies need not
stay constant (such as in a potential with multiple pattern frequencies
from multiple causes).

The Python package presented here, \texttt{commensurability}, provides a
straightforward Python framework and connection to powerful tools for
modeling potentials to estimate the volume of a toroid that a given
orbit fills.

\section{Statement of need}\label{statement-of-need}

The technique of measuring orbit volumes has previously been used to
classify orbits in 2D in both fixed and evolving potentials, revealing
families of commensurate orbits that are critical for galactic evolution
(\citeproc{ref-Petersen.Weinberg.Katz.2021}{Petersen et al., 2021}).
However, the technique had not (1) been extended to 3D, and (2) did not
have a user-friendly Python interface. The \texttt{commensurability}
package solves both problems and provides interoperability with three
leading galactic dynamics packages to accomplish the underlying orbit
integrations.

More generally, classifying orbits in model galaxies, let alone in
models that evolve with time, is a non-trivial task. The primary method
to classify orbits in galaxies relies on frequency analysis, such as
with the SuperFreq (\citeproc{ref-superfreq.paper}{Price-Whelan, 2015})
or \texttt{naif} (\citeproc{ref-naif.paper}{Silva et al., 2023}) codes.
While frequency analysis is a useful tool, there remains ambiguity in
the classification of commensurate orbits over short durations, or while
frequencies change. Frequency analysis relies on the stability of the
frequencies of an orbit, which may miss short-lived families and
smoothly changing frequencies. Orbit tessellation avoids these pitfalls
by estimating the volume of the toroid the orbit would traverse in the
long-time limit, even if frequencies are changing over time. Orbit
tessellation improves on fundamental frequency classifications, but also
exists as an independent orbit classification scheme that can operate on
orbits run in self-consistent simulations.

Additionally, computing the orbit volume provides an opportunity to
efficiently search the model potential space for distinct orbital
families. By creating a measure that is defined at all points in phase
space (typically defined in the position and velocity of a radial
extreme), one can search a model potential for commensurate orbits using
optimization techniques.

\begin{figure}
\centering
\includegraphics{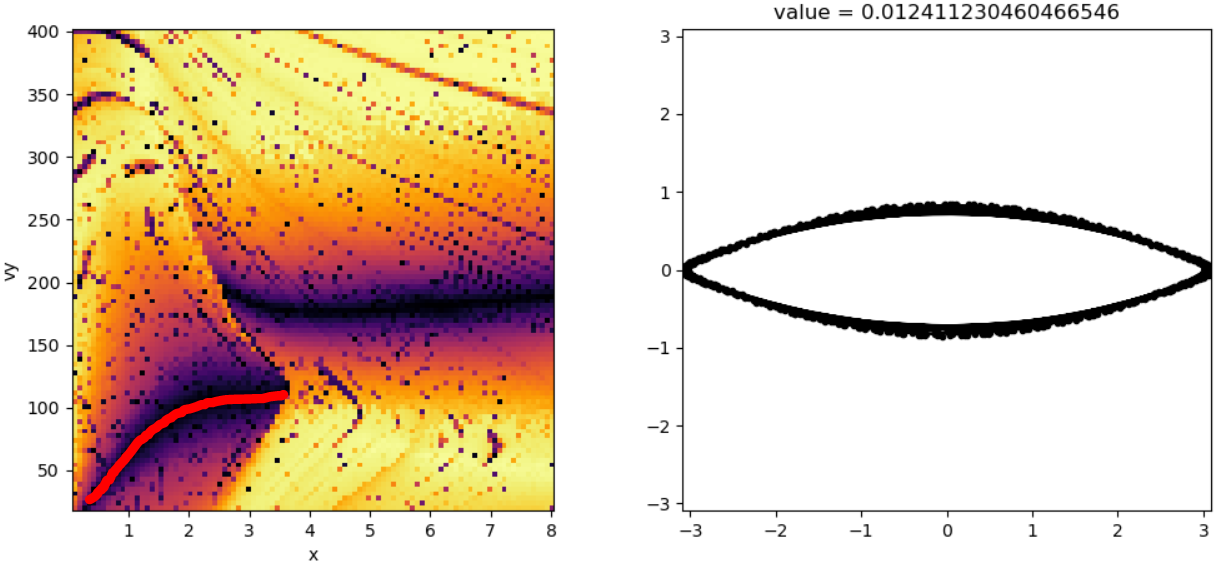}
\caption{An example diagnostic from commensurability, where an orbit is
selected from a map of toroid volumes in radial position \(x\) and
tangential velocity \(v_y\) (left panel). In the map, orbits with small
measures are dark-colored. The ``tracks'', including the curve
highlighted in red, are families of orbits. The orbits are all in the
frame of a rotating bar; the shape traced by this particular orbit in
x-y position space (right panel) is a characteristic bar orbit, drawn
from a point on the red curve. \label{fig:x1orbit}}
\end{figure}

\section{Features and dependencies}\label{features-and-dependencies}

\texttt{commensurability} provides several notable features, with an
emphasis on extensibility and compatibility with existing code.

\texttt{commensurability} defines a framework for analyzing potentials
in its ``analysis'' objects. ``TessellationAnalysis'' uses
\href{https://docs.python.org/3/library/multiprocessing.html}{multiprocessing}
to efficiently compute the normalized toroidal volumes of orbits, or
evaluate orbit ``measures''. Analysis objects are equipped with
interactive Matplotlib (\citeproc{ref-matplotlib.paper}{Hunter, 2007})
figures displaying a map of orbit measures as a function of phase space
(for instance in position-velocity \(x,v_y\); \autoref{fig:x1orbit}). A
user can explore the potential by visually selecting regions of
interest, and the interactive plot will update with the \(x,y,z\) time
series of the orbit. Since evaluating a large number of orbits can
require significant computational power, analysis objects also include
serialization and deserialization methods using the HDF5 format
(\citeproc{ref-hdf5.paper}{The HDF Group, n.d.}). Although
TessellationAnalysis objects focus on orbit tessellation specifically,
the commensurability analysis framework can be easily extended to any
orbit evaluation method using the abstract base class AnalysisBase.

\texttt{commensurability} provides a subpackage called tessellation that
implements mathematical and computational improvements over early
versions of orbit tessellation-based classification
(\citeproc{ref-Petersen.Weinberg.Katz.2021}{Petersen et al., 2021}).
While only two- and three-dimensional orbits are evaluated in
\texttt{commensurability}, the subpackage implements a general
N-dimensional evaluation algorithm. Several normalizations are provided
for the toroid volume, computed after trimming insignificant simplices
based on their axis ratio, and new normalizations can be defined easily.
The default normalization in 3D for the orbit volume is the convex hull
of four rotated copies of the points around the \(z\)-axis (the rotation
axis of the model galaxy). The rotated copies protect against missing
orbits that only span a small range of azimuth about the \(z\)-axis. The
convex hull is computed using QHull (\citeproc{ref-qhull.paper}{Barber
et al., 1996}) as implemented in SciPy
(\citeproc{ref-scipy.paper}{Virtanen et al., 2020}). This subpackage can
be used for its orbit evaluation function independent of
commensurability.

\texttt{commensurability} depends on the Python package
\href{https://github.com/ilikecubesnstuff/pidgey}{pidgey} for orbit
integration. Pidgey is a standalone package that asserts a uniform
interface to three major galactic dynamics packages---AGAMA
(\citeproc{ref-agama.paper}{Vasiliev, 2019}), \texttt{gala}
(\citeproc{ref-gala.paper}{Price-Whelan, 2017}), and galpy
(\citeproc{ref-galpy.paper}{Bovy, 2015})---and its interface can be
extended to more packages trivially. Pidgey uses Astropy
(\citeproc{ref-astropy.paper}{Astropy Collaboration et al., 2022})
SkyCoord objects to store orbit coordinates, a format familiar to most
astronomers. Pidgey depends on
\href{https://github.com/ilikecubesnstuff/iext}{iext}, a defensive
framework for extending the dependencies associated with a class without
requiring the dependencies to be present. This enables pidgey to operate
in the absence of a complete suite of orbit integration packages; the
user need only have one of AGAMA, \texttt{gala}, or galpy installed to
use commensurability (but may have all three).

\section{Acknowledgements}\label{acknowledgements}

We thank Aneesh Naik and Eugene Vasiliev for discussions regarding early
versions of the software.

Funded by the European Union (MSCA EDUCADO, GA 101119830). Views and
opinions expressed are however those of the author(s) only and do not
necessarily reflect those of the European Union. Neither the European
Union nor the granting authority can be held responsible for them.

\section*{References}\label{references}
\addcontentsline{toc}{section}{References}

\phantomsection\label{refs}

\end{document}